\documentclass[twocolumn,10pt,aps,pra,a4paper,amsmath,amssymb,floatfix]{revtex4-1}
\pdfoutput=1 
\usepackage[utf8]{inputenc}
\usepackage{amsmath}
\usepackage{graphicx}   
\usepackage{dcolumn}    
\usepackage{bm}         

\newcommand{\Cfsnv}[1]{C_{\rm fs}^{[N]}(#1)}

\newcommand{\Cfsxv}[1]{C_{\rm fs}^{[\bm{\Xi}]}(#1)}

\newcommand{\Cfsun}[1]{C_{\rm fs \neq}^{[\bm{\Xi}]}(#1)}
\newcommand{\eb}{e^{-\beta \hat H}}

\newcommand{\ebN}{e^{-\beta_N \hat H}}
\newcommand{\ebt}{e^{-\beta \hat H/2}}

\newcommand{\etf}{e^{-i \hat H t/\hbar}}
\newcommand{\etb}{e^{i \hat H t/\hbar}}
\newcommand{\etfo}{e^{-i \hat H_0 t/\hbar}}
\newcommand{\etbo}{e^{i \hat H_0 t/\hbar}}

\newcommand{\bra}[1]{\langle #1 |}
\newcommand{\ket}[1]{| #1 \rangle}
\newcommand{\kb}[1]{\ket{#1}\bra{#1}}

\newcommand{\dfs}[1]{\delta\!\left[ #1 \right]}
\newcommand{\dfQ}{\delta[f({\bf Q})]}

\newcommand{\ddp}[2]{\frac{\partial #1}{\partial #2}}

\newcommand{\piNz}{\prod_{i=0}^{N-1}}

\newcommand{\smiN}{\sum_{i=1}^N}

\newcommand{\smiNz}{\sum_{i=0}^{N-1}}
\newcommand{\lttz}{\lim_{t\to 0_+}}

\newcommand{\shortt}{$t\to 0_+$}
\newcommand{\longt}{$t\to \infty$}
\newcommand{\largeN}{$N\to\infty$}

\newcommand{\tr}{ {\rm Tr} }

\newcommand{\td}{\tilde}
\newcommand{\tdo}{\tilde D_0}
\newcommand{\tpo}{\tilde P_0}
\newcommand{\Qrb}{Q_{\rm r}(\beta)}
\newcommand{\eqr}[1]{Eq.~\eqref{eq:#1}}
\newcommand{\ort}{\tfrac{1}{\sqrt{2}}}
\newcommand{\otrt}{\tfrac{1}{2\sqrt{2}}}

\newcommand{\ntn}{\mathcal{N}_{2N}}
\newcommand{\no}{\nonumber}

\newcommand{\bq}{ {\bf q} }

\newcommand{\by}{ {\bf y} }
\newcommand{\bz}{ {\bf z} }
\newcommand{\bDelta}{ {\bf \Delta} }
\newcommand{\bmeta}{ \bm{\eta} }
\newcommand{\bQ}{ {\bf Q} }
\newcommand{\bZ}{ {\bf Z} }
\newcommand{\bP}{ {\bf P} }
\newcommand{\bD}{ {\bf D} }
\newcommand{\fq}{ f({\bf q}) }
\newcommand{\fQ}{ f({\bf Q}) }

\newcommand{\fz}{ f({\bf z}) }

\newcommand{\gz}{ g({\bf z}) }
\newcommand{\gq}{ g({\bf q}) }
\newcommand{\gQ}{ g({\bf Q}) }

\newcommand{\tphN}{\frac{1}{(2\pi\hbar)^N}}
\newcommand{\tphtN}{\frac{1}{(2\pi\hbar)^{2N}}}
\newcommand{\inti}{\int_{-\infty}^{\infty}}
 
\begin{document}
\bibliographystyle{/home/tjhh2/Projects/lib/latexabbrev/tim}

\title{On the uniqueness of \shortt~quantum transition-state theory} 
\author{Timothy J.~H.~Hele}
\email{tjhh2@cam.ac.uk}
\author{Stuart C.~Althorpe}
\affiliation{Department of Chemistry, University of Cambridge, Lensfield Road, Cambridge, CB2 1EW, UK.}
\date{\today}

\begin{abstract}
It was shown recently that there exists a true quantum transition-state theory (QTST) corresponding to the \shortt~limit of a (new form of)
 quantum flux-side time-correlation function. Remarkably, this QTST is {\em identical} to ring-polymer molecular dynamics (RPMD) TST. Here we provide evidence which suggests very strongly that this QTST ($\equiv$ RPMD-TST) is unique, in the sense that the \shortt~limit of any other flux-side time-correlation function gives either non-positive-definite quantum statistics or zero. We introduce a generalized flux-side time-correlation function which includes {\em all} other (known) flux-side time-correlation functions as special limiting cases. We find that the only non-zero \shortt~limit of this function that contains
 positive-definite quantum statistics is RPMD-TST.
 \emph{Copyright (2013) American Institute of Physics. This article may be downloaded for personal use only. Any other use requires prior permission of the author and the American Institute of Physics. The following article appeared in The Journal of Chemical Physics, \textbf{139} (2013), 084116, and may be found at http://link.aip.org/link/?JCP/139/084116/1}
\end{abstract}

\maketitle

\section{Introduction}
\label{sec:intro}
Classical transition-state theory has enjoyed wide applicability and success in calculating the rates of chemical processes \cite{cha78, fre02, eyr35,tru96}. Its central premise\cite{eyr35rev} is the assumption that all trajectories which cross the barrier react (rather than recross)\footnote{This article is concerned with configuration-space TST, not the formally exact phase-space TST as discussed in S.~Wiggins, L.~Wiesenfeld, C.~Jaff\'e and T.~Uzer, \emph{Phys. Rev. Lett.}  \textbf{86} (2001), 5478.}. This was subsequently recognized as being equivalent to taking the short-time limit of a classical flux-side time-correlation function \cite{cha78,fre02}, whose long-time limit would be the exact classical rate \cite{mil74}.

Until very recently it was thought that there was no rigorous quantum generalization of classical transition-state theory\cite{vot93,mil93,pol05}, because the \shortt~limit of all known quantum flux-side time-correlation functions was zero, i.e.~there was no short-time quantum rate theory which would produce the exact rate in the absence of recrossing. Nevertheless, a large variety of `Quantum Transition-State Theories' (QTSTs) have been proposed \cite{gil87,vot89rig,cal77,ben94,pol98,and09,vot93,tru96,gev01,shi02,pol98,mci13} using heuristic arguments, along with other methods of obtaining the reaction rate from short-time data \cite{mil03,van05,wan98,wan00,rom11,rom11adapt,wol87}.

However, in two recent papers \cite{hel13,alt13} (hereinafter Paper I and Paper II) we showed that a vanishing \shortt~limit arises only because the standard forms of flux-side time-correlation function use flux and side dividing surfaces that are different functions of (imaginary-time) path-integral space. When the flux and side dividing surfaces are chosen to be the same, the \shortt~limit becomes non-zero.

Initially, we thought that there would be many types of computationally useful \shortt~quantum TST, since there is an infinite number of ways in which one can choose a common dividing surface in path integral space. For example, one can choose the surface to be a function of just a single point (in path-integral space), in which case one recovers at \shortt~the simple form of quantum TST that was introduced on heuristic grounds by Wigner\cite{wig32uber,mil75} (and used to obtain his famous expression for parabolic-barrier tunnelling). However, this form of TST becomes negative at low temperatures,\cite{hel13, liu09, and09} because the single-point dividing surface constrains the quantum Boltzmann operator in a way that makes it non-positive-definite. To obtain positive-definite quantum statistics, it is necessary to choose dividing surfaces that are invariant under cyclic permutation of the polymer beads, since this preserves imaginary-time translation in the infinite-bead limit. Under this strict condition, the \shortt~limit is guaranteed to be positive definite and, remarkably, is identical to ring-polymer molecular dynamics TST (RPMD-TST).

This last result is useful because it shows that the powerful techniques of RPMD rate theory\cite{man05che, man04, man05ref, hab13, men11, kre13, ric13, per12, lix13,col09,col10,sul11,sul12} and the earlier-derived centroid TST\cite{gil87, vot89rig} are not heuristic guesses (as was previously thought), but are instead rigorous calculations of the instantaneous thermal quantum flux from reactants to products.\footnote{Note that direct application of
 RPMD rate theory [i.e. exact classical rate theory applied in the extended (fictitious) ring-polymer space] gives a lower bound to the RPMD-TST result, and thus a lower bound to the instantaneous quantum flux through the ring-polymer dividing surface.}

The quantum TST referred to above (i.e. RPMD-TST) is unique, in the sense that any other type of dividing surface gives non-positive-definite quantum statistics, when introduced into the ring-polymerised flux-side time-correlation function that was introduced in Paper~I. However, the question then arises as to whether there are \shortt~limits of {\em different} flux-side time-correlation functions, which also give positive-definite quantum statistics, but which are different from (and perhaps better than!) RPMD-TST. Here we give very strong evidence (though not a proof) that this is not the case, and that RPMD-TST is indeed the unique \shortt~quantum TST.

After summarizing previous work in Sec~\ref{sec:rev}, we write out in Sec~\ref{sec:gqfs} the most general form of quantum flux-side dividing surface that we have been able to devise. We cannot of course prove that a more general form does not exist, but we find that the new correlation function is sufficiently general that it includes {\em all} other known flux-side time-correlation functions as special cases. In Sec~\ref{sec:stl}, we take the \shortt~limit of this function and obtain a set of conditions which are necessary and sufficient for the \shortt~limit to be non-zero and positive-definite. We find that these conditions give RPMD-TST. Section \ref{sec:cc} concludes the article.

\section{Review of earlier developments}
\label{sec:rev}
To simplify the algebra, the following is presented for a one-dimensional system with coordinate $x$, mass $m$ and Hamiltonian $\hat H$ at an inverse temperature $\beta \equiv 1/k_B T$. The results generalize immediately to multi-dimensional systems, as discussed in Paper I.
We begin with the Miller-Schwarz-Tromp (MST) expression for the exact quantum mechanical rate \cite{mil74,mil83},
\begin{align}
 k^{\rm QM}(\beta) = \lim_{t\to \infty}c_{\rm fs}^{\rm sym}(t)/\Qrb,
\end{align}
where $\Qrb$ is the reactant partition function, and
\begin{align}
 c_{\rm fs}^{\rm sym}(t) = \tr \left[ \ebt \hat F \ebt \etb \hat h \etf \right]
\label{eq:milsym}
\end{align}
where $\hat F$ is the quantum-mechanical flux operator
\begin{align}
 \hat F = \frac{1}{2m}\left[\delta(x - q^\ddag) \hat p + \hat p \delta(x - q^\ddag) \right]
\end{align}
and $\hat h$ is the heaviside operator projecting onto states in the product region, defined relative to the dividing surface $q^\ddag$.

The function $c_{\rm fs}^{\rm sym}(t)$ tends smoothly to zero in the \shortt~limit, \cite{vot89time,vot93,mil93} which would seem to rule out the existence of a  \shortt \ quantum transition-state theory. However, 
it was shown in Paper~I that this behaviour arises because the flux and side dividing surfaces in \eqr{milsym} are {\em different} functions of path-integral space \cite{hel13}. When the two dividing surfaces are the same, the quantum flux-side time-correlation function becomes non-zero in the \shortt \ limit. 
(Note that the classical flux-side time-correlation function also tends smoothly to zero as \shortt \ if the flux and side dividing surfaces are different.) A simple form of quantum
flux-side time-correlation function in which the two surfaces are the same is 
\begin{align}
 C_{\rm fs}^{[1]}(t) = & \int dq \int dz \int d \Delta \ h(z) \hat F(q) \nonumber \\
 & \times \bra{q - \Delta/2} \eb \ket{q+\Delta/2} \no \\ 
 & \times \bra{q+\Delta/2} \etb \kb{z} \etf \ket{q-\Delta/2}.
\label{eq:ubf}
\end{align}
where the superscript $[1]$ indicates that the common dividing surface is a function of a single-point in path integral space. In the \longt~limit, \eqr{ubf} gives the exact quantum rate. In the \shortt~limit, \eqr{ubf} is non-zero (because the dividing surfaces are the same), and thus gives a \shortt~QTST, which is found to be identical to one proposed on heuristic grounds by Wigner in 1932 \cite{wig32uber} and later by Miller \cite{mil75}. Unfortunately, this form of QTST becomes negative at low temperatures, because the constrained quantum-Boltzmann operator is not positive-definite, and thus gives an erroneous description of the quantum statistics \cite{liu09,hel13,and09}. 

Paper~I showed that positive-definite quantum statistics can be obtained using a ring-polymerized flux-side time-correlation function of the form
\begin{align}
 \Cfsnv{t} = & \int d\bq \int d{\bf \Delta} \int d\bz \ \mathcal{\hat F}[\fq] h[\fz] \nonumber \\
& \times \prod_{i=0}^{N-1} \bra{q_{i-1}-\tfrac{1}{2}\Delta_{i-1}} \ebN \ket{q_i+\tfrac{1}{2}\Delta_i} \no \\
& \qquad \times \bra{q_i+\tfrac{1}{2}\Delta_i} \etb \ket{z_i} \nonumber \\
& \qquad \times \bra{z_i} \etf \ket{q_{i}-\tfrac{1}{2}\Delta_{i}}, \label{eq:ubfn}
\end{align}
where the integrals extend over the whole of path-integral space ($\int d\bq \equiv \inti dq_0 \ldots \inti dq_{N-1}$ and so on),  and $\fq$ is the common dividing surface, which is chosen to be invariant under cyclic permutation of the arguments $\bq$ or $\bz$. The `ring-polymer flux operator' $\mathcal{\hat F}[\fq]$ describes the flux perpendicular to $\fq$, and is given by
\begin{align}
 \mathcal{\hat F}[\fq] = & \frac{1}{2m} \smiNz \Bigg\{\ddp{\fq}{q_i} \dfs{\fq} \hat p_i \no\\
 & \qquad + \hat p_i \dfs{\fq} \ddp{\fq}{q_i}\Bigg\} \label{eq:sfo} 
\end{align}
where the first term in braces is placed between $\ebN \ket{q_i+\tfrac{1}{2}\Delta_i}$ and $\bra{q_i+\tfrac{1}{2}\Delta_i} \etb$, and the second term between  $\etf \ket{q_{i}-\tfrac{1}{2}\Delta_{i}}$ and $\bra{q_{i}-\tfrac{1}{2}\Delta_{i}} \ebN$. \footnote{See Sec.~IV~B of Paper~I (Ref.~\cite{hel13}) for details.} We then take the limits
\begin{align}
  \lttz & \lim_{N\to\infty}\Cfsnv{t} = \no \\
  & \int d{\bf Q} \ \dfQ \sqrt{\frac{{\cal N}_N}{2\pi m \beta}} \piNz \bra{Q_{j-1}} e^{-\beta_N \hat H} \ket{Q_j} \nonumber\\
  & = k_{Q}^\ddag(\beta)\Qrb,
  \label {eq:nopeq} 
\end{align}
where
\begin{align}
 {\cal N}_N  
  & = N \smiNz \left[ \ddp{\fQ}{Q_i} \right]^2 
\end{align}
and $k_{Q}^\ddag(\beta)$ is the quantum TST rate,  which is guaranteed to be positive, because the cyclic-permutational invariance of $\fq$ ensures that the constrained Boltzmann operator is positive-definite. Unlike \eqr{ubf}, \eqr{ubfn} does not give the exact quantum rate in the limit \longt. However, we showed in Paper II that \eqr{ubfn}
does give the exact quantum rate if there is no recrossing of the dividing surface $\fq$, and thus that $k_{Q}^\ddag(\beta)$ is a good approximation to the exact quantum rate if the amount of such recrossing is small.

 Remarkably, 
\begin{align}
 k_{Q}^\ddag(\beta) \equiv k_{\rm RPMD-TST}^\ddag(\beta) 
\end{align}
where $k_{\rm RPMD-TST}^\ddag(\beta)$ is the ring-polymer molecular dynamics TST (RPMD-TST) rate, corresponding to the \shortt~limit of the (classical) flux-side time-correlation function in ring-polymer space.  Hence \eqr{ubfn} gives a rigorous justification of the powerful method of RPMD-TST (and also of centroid-TST), by showing that it is a computation of the short time quantum flux (rather than merely an heuristic approach, as was previously thought\cite{man05che, ric09, alt11}).

As mentioned above, the dividing surface $\fq$ is invariant under cyclic permutation of the coordinates $\bq$ and $\bz$, meaning that $\fq$ is invariant under imaginary-time translation in the limit \largeN. In Paper I, we showed that only if this condition is met does the \shortt~limit of \eqr{ubfn} give positive-definite quantum statistics in the limit \largeN. Hence, if we start with the flux-side time-correlation function \eqr{ubfn}, the quantum TST rate $ k_{Q}^\ddag(\beta)\equiv k_{\rm RPMD-TST}^\ddag(\beta)$ of \eqr{ubfn} is unique, in the sense that any other \shortt~limit [i.e. using a non-cyclically invariant $\fq$] does not give positive-definite quantum statistics. 

\section{General quantum flux-side time-correlation function}
\label{sec:gqfs}

\begin{table*}[tbh]
\begin{tabular}{l|c|c|c|c|c|c}
 Flux-side t.c.f.			& $N$ 	& $\xi_i^-$ 	& $\xi_i^+$ 	& $\mathcal{\hat F}[\fq]$	& $h[\gz]$ & \shortt~limit \\ \hline
 Miller-Schwarz-Tromp \cite{mil83} 			& 2 	& 1/2		& 0 		& $\mathcal{\hat F}(q_1)$ 	& $h(z_0)$ & 0 \\
 Asymmetric MST \cite{mil83}			& 2	& $\xi_1^- = 1, \ \xi_2^- = 0$& 0 & $\mathcal{\hat F}(q_1)$	& $h(z_0)$ & 0  \\
 Kubo-transformed \cite{man05che}				&$\infty$& $1/N$	& 0 		& $\mathcal{\hat F}(q_0)$	& $\sum_{i=1}^{N-1} h(z_i)$ & 0  \\
 Wigner [$C_{\rm fs}^{[1]}(t) $ of Eq.~\eqref{eq:ubf}]			& 1	& 1		& 0		& $\mathcal{\hat F}(q_0)$	& $h(z_0)$ & Wigner TST \cite{wig32uber}  \\
 $C_{\rm fs}^{[1]}(t)'$ of Eq.~\eqref{eq:opff}		& 1	& 1/2		& 1/2		& $\mathcal{\hat F}(q_0)$	& $h(z_0)$ & Double-Wigner TST  \\
 Hybrid [Eq.~7 of Ref.~\onlinecite{alt13}]					& $>\!1$	& $1/N$		& 0		& $\mathcal{\hat F}[\fq]$	& $h(z_0)$ & 0  \\
Ring-polymer [$\Cfsnv{t}$ of Eq.~\eqref{eq:ubfn}]			& $\infty$	& $1/N$		& 0		& $\mathcal{\hat F}[\fq]$	& $h[\fz]$ & RPMD-TST  \\
 \end{tabular}
 
\caption{How to generate every (known) form of flux-side time-correlation function as a special case of Eq.~\eqref{eq:gencfs}.
 The terms $\xi_i^-$,  $\xi_i^+$,  $\mathcal{\hat F}[\fq]$ and $h[\gz]$ are defined in Eq.~\eqref{eq:gencfs}. Double-Wigner TST is the
  generalization of Wigner-TST that results from the \shortt~limit of Eq.~\eqref{eq:ubf}. In the hybrid and ring-polymer expressions, $\fq$ is chosen to be invariant under cyclic permutation of the coordinates $q_i$; RPMD-TST specialises to centroid-TST when $\fq=\sum_{i=0}^{N-1} q_i/N$.} 
\label{tab:cfs}
\end{table*}

The question then arises as to whether other QTSTs exist, obtained by taking the \shortt~limit of other flux-side time-correlation functions, which also give positive-definite quantum statistics.  It is clear that \eqr{ubfn} is not the most general flux-side time-correlation function with such a limit because one can modify \eqr{ubf} to give a `split Wigner flux-side time-correlation function':
\begin{align}
 {C_{\rm fs}^{[1]}}'(t) = & \int dq \int dz \int d \Delta \int d \eta \ h(z) \mathcal{\hat F}(q) \nonumber \\
 & \times \bra{q - \Delta/2} \ebt \ket{q+\Delta/2} \no\\
 & \times \bra{q+\Delta/2}\etb \ket{z-\eta/2} \no \\
 & \times \bra{z-\eta/2} \ebt \ket{z+\eta/2} \no \\
 & \times \bra{z+\eta/2} \etf \ket{q-\Delta/2}.
\label{eq:opff} 
\end{align}
which is easily shown to give the exact quantum rate in the \longt~limit and to have a non-zero \shortt~limit. This limit is not positive-definite, but clearly one could imagine generalizing \eqr{opff} in the analogous way to which eq \eqr{ubfn} is obtained by ring-polymerizing \eqr{ubf}. 

A form of flux-side time-correlation function which does include \eqr{opff}, as well as a ring-polymerized generalization of it, is 
\begin{align}
 \Cfsun{t} = & \int d\bq \int d\bz \int d\bDelta \int d\bmeta\ \mathcal{\hat F}[\fq]h[\gz] \nonumber \\
 \times & \piNz \bra{q_{i-1}-\Delta_{i-1}/2} e^{-\beta \xi_i^- \hat H} \ket{q_i+\Delta_i/2} \no \\
 & \times \bra{q_i+\Delta_i/2} e^{i\hat H t /\hbar} \ket{z_i - \eta_i/2} \nonumber \\
 & \times \bra{z_i - \eta_i/2} e^{-\beta \xi_i^+ \hat H} \ket{z_i+\eta_i/2} \no \\ 
 & \times \bra{z_i+\eta_i/2} e^{-i\hat H t /\hbar} \ket{q_i - \Delta_i/2}.
 \label{eq:gencfs} 
\end{align}
Here the imaginary time-evolution has been divided into pieces of varying lengths $\xi_i^\pm\beta\hbar$, which are interspersed with forward-backward real-time propagators. To set the inverse temperature $\beta$, we impose the requirement
\begin{align}
 \smiNz \xi_i^- + \xi_i^+ = 1, \label{eq:sumtoone}
\end{align}
where $\xi_i^{\pm} \ge 0\ \forall i$. 
The only restrictions, at present, on the dividing surface $\fq$ are  \begin{align}
 \lim_{q\to\infty} f(q,q,\ldots,q) > 0, \\
 \lim_{q\to -\infty} f(q,q,\ldots,q) < 0.
\end{align}
and similarly for $\gq$. [These are simply the conditions that are necessary for $\fq$ and $\gq$ to distinguish reactants from products and thus do their jobs as dividing surfaces.] The subscript $\neq$ symbolises that the dividing surfaces are not necessarily equal.
Equation~\eqref{eq:gencfs} is represented diagrammatically in Fig.~1a.

The function  $\Cfsun{t}$ correlates the flux averaged over a set of imaginary-time paths with the side averaged over another set of imaginary-time paths at some later time $t$.
Every form of quantum flux-side time-correlation function (known to us) can be obtained either directly from $\Cfsun{t}$, using particular choices of $\fq$, $\gq$ and $\bm{\xi}$, or or by taking linear combinations of $\Cfsun{t}$ containing different values of these parameters; see Table~\ref{tab:cfs}. We believe that $\Cfsun{t}$ is the most general expression yet obtained for a quantum flux-side time-correlation function (before taking linear combinations), although we cannot prove that a more general expression does not exist. 

\section{The short-time limit}
\label{sec:stl}

We now take the \shortt~limit of \eqr{gencfs}, and determine the conditions under which this limit is non-zero and contains positive-definite quantum statistics.\footnote{The derivation presented here assumes a single potential energy surface. It could also be performed for a system with multiple potential energy surfaces by treating the potential in the manner of M.~H~Alexander, \textit{Chem.~Phys.~Lett.} \textbf{347} (2001), 436.}

\subsection{Non-zero \shortt~limit}
In order to calculate the short-time limit of \eqr{gencfs} we first note that 
\begin{align}
 \lim_{t\to 0_+} & \bra{x} \etb \kb{y} \etf \ket{z} \no \\
 & = \bra{x} \etbo \kb{y} \etfo \ket{z}
 \label{eq:shorttlim}
\end{align}
where $\hat H_0 = \hat p^2/2m$ is the free particle Hamiltonian, and that
\begin{align}
 \bra{x} \etfo \ket{y} = & \sqrt{\frac{m}{2\pi i \hbar t}} e^{im(x-y)^2/2\hbar t} \label{eq:etfo} \\
 \bra{x} \etfo \hat p \ket{y} = & \frac{(x-y)m}{t}\sqrt{\frac{m}{2\pi i \hbar t}} e^{im(x-y)^2/2\hbar t}. \label{eq:etfop}
\end{align}
We then substitute the identity
\begin{align}
 e^{-\beta \xi_i^+ \hat H} \equiv & \int dy_i \int d \zeta_i \ \etf \ket{y_i - \zeta_i/2} \no\\
 & \times \bra{y_i - \zeta_i/2} e^{-\beta \xi_i^+ \hat H} \ket{y_i + \zeta_i/2} \no \\
 & \times \bra{y_i + \zeta_i/2} \etb.
\end{align}
 into \eqr{gencfs}, to obtain
\begin{align}
  \Cfsun{ & t\to 0_+} = \no \\
  & \lttz \int d\bq \int d\bz \int d\bDelta \int d\bmeta \int d\by \int d\bm{\zeta} \no \\
  & \times \ \mathcal{\hat F}[\fq]h[\gz] \nonumber \\
 & \times \piNz \bra{q_{i-1}-\Delta_{i-1}/2} e^{-\beta \xi_i^- \hat H} \ket{q_i+\Delta_i/2} \no \\
 & \qquad \times \bra{q_i+\Delta_i/2}\etb \ket{z_i - \eta_i/2} \no \\
 & \qquad \times \bra{z_i - \eta_i/2} \etf \ket{y_i - \zeta_i/2} \nonumber \\
 & \qquad \times \bra{y_i - \zeta_i/2} e^{-\beta \xi_i^+ \hat H} \ket{y_i + \zeta_i/2} \no \\
 & \qquad \times \bra{y_i + \zeta_i/2} \etb \ket{z_i+\eta_i/2} \no \\
 & \qquad \times \bra{z_i+\eta_i/2} \etf \ket{q_i - \Delta_i/2} \label{eq:nfs}. 
\end{align}

The imaginary-time propagators in \eqr{nfs} alternate with pairs of forward-backward real-time propagators, which allows us to use Eqs.~\eqref{eq:shorttlim}--\eqref{eq:etfop} to take the \shortt~limit\footnote{One can evaluate the short-time limit of {\protect \eqr{gencfs}} without the insertion of the unit operators, but their use greatly simplifies the subsequent algebra.}. This procedure is straightforward, but algebraically lengthy, so we give only the main steps here, relegating the details to Appendix~\ref{ap:st}.

The first step (Sec.~\ref{ssec:cot}) is to transform \eqr{nfs} to 
\begin{align}
  \Cfsun{t} = & \int d{\bf Q} \int d {\bf Z} \int d {\bf D} \ \mathcal{\hat F}[f({\bf Q,D})]h[g({\bf Z})] \nonumber \\
 \times & \prod_{j=0}^{2N-1} \bra{Q_{j-1}-D_{j-1}/2} e^{-\beta \xi_j \hat H} \ket{Q_j+D_j/2} \no \\
 \times & \bra{Q_j+D_j/2} \etb \ket{Z_j} \nonumber \\
 \times & \bra{Z_j} \etf \ket{Q_j - D_j/2}
 \label{eq:simpt} 
\end{align}
where $\bQ\equiv\{Q_j\}$, $j=0\dots2N-1$, and similarly for $\bZ$, $\bD$, 
and
\begin{align}
\xi_{2i}&=\xi^-_i\\
\xi_{2i+1}&=\xi^+_i,\ \ i=0,\dots,N-1
\end{align}
 i.e.\
we have halved the number of bra-kets in each imaginary time-slice, by doubling the number of polymer beads. Equation~\eqref{eq:simpt} is superficially similar to Eq.~31 of Paper I, but differs from it in the important respect that the dividing surface $\fq$ now depends on the coordinate $\bD$ (in the way described in Sec.~\ref{ssec:cot}). As a result the flux and side dividing surfaces are in general {\em different} functions of path integral space, even if we choose $\fq \equiv \gq$. 
\clearpage 
On the basis of Paper I, one might therefore expect the \shortt~limit of \eqr{simpt} to be zero, except for the special cases corresponding to Wigner TST and RPMD-TST  (given in Table I). However, we show in Sec.~A2 that the \shortt~limit of \eqr{simpt} is {\em always} non-zero when $\fq \equiv \gq$, because the $\bD$-dependence of $f({\bf Q,D})$ integrates out in this limit, to give

\begin{widetext}
\begin{align}
\lttz\Cfsxv{t} = & \frac{1}{(2\pi\hbar)^N} \int d{\bf Q} \int d {\bf P^+} \int d {\bf D^+} \ \dfQ S_f({\bf Q,P^+})h[S_f({\bf Q,P^+})]  \nonumber \\
  \times &  \prod_{i=0}^{N-1} \bra{Q_{2i-1}-\otrt D_{i-1}^+} e^{-\beta \xi_j \hat H} \ket{Q_{2i}+\otrt D_i^+} \bra{Q_{2i}-\otrt D_i^+} e^{-\beta \xi_j \hat H} \ket{Q_{2i+1}+\otrt D_i^+} e^{i D^+_j P^+_j/\hbar} \label{eq:nzq} 
\end{align}
\end{widetext}
where $\bP^+$ and $\bD^+$ are the $N$-dimensional vectors defined in Sec~\ref{ssec:stl}, $S_f({\bf Q,P^+})$ is the flux perpendicular to $\fQ$, and the absence of a subscript $\neq$ in $\Cfsxv{t}$ indicates $\fq \equiv \gq$. Thus, in general, $f({\bf Q,D})$ acts as a time-dependent flux-dividing surface, which becomes the same as the side-dividing surface in the limit \shortt~if $\fq \equiv \gq$. Clearly $\fq$ is time-independent in the special case that $\xi_i^-=1/N,\xi_i^+=0$, in which \eqr{gencfs} reduces to \eqr{ubfn} (see Table I).

We can tidy up \eqr{nzq} by integrating out $(N-1)$ of the integrals in $\bP^+$ and $\bD^+$ (see Sec.~\ref{ssec:intout}), to obtain
\begin{align}
 & \lttz \Cfsxv{t} = \frac{1}{2\pi\hbar} \int d{\bf Q} \int d \tilde P_0 \int d \tilde D_0 \no \\ 
 & \times h[\tilde P_0]\frac{\tilde P_0}{m}  \sqrt{B_N} \dfQ e^{i \tilde D_0 \tilde P_0/\hbar} \nonumber \\
 & \times \prod_{j=0}^{2N-1} \bra{Q_{j-1}- T_{j-1\ 0} \tilde D_0/2} e^{-\beta \xi_j \hat H} \ket{Q_j+ T_{j0} \tilde D_0/2}.
\label{eq:likenf2} 
\end{align}
where $\tpo$ is the momentum perpendicular to the dividing surface $\fQ$, 
 $\tdo$ describes a collective ring-opening mode, $T_{j0}$ is the weighting of the $j$th path-integral bead in the dividing surface $\fQ$ [see Eq.~\eqref{eq:tdef}],
 and $\sqrt{B_N}$ is a normalization constant associated with $\tpo$.
  
\begin{figure*}[tbh]
\resizebox{.8\textwidth}{!}{\includegraphics[angle=270]{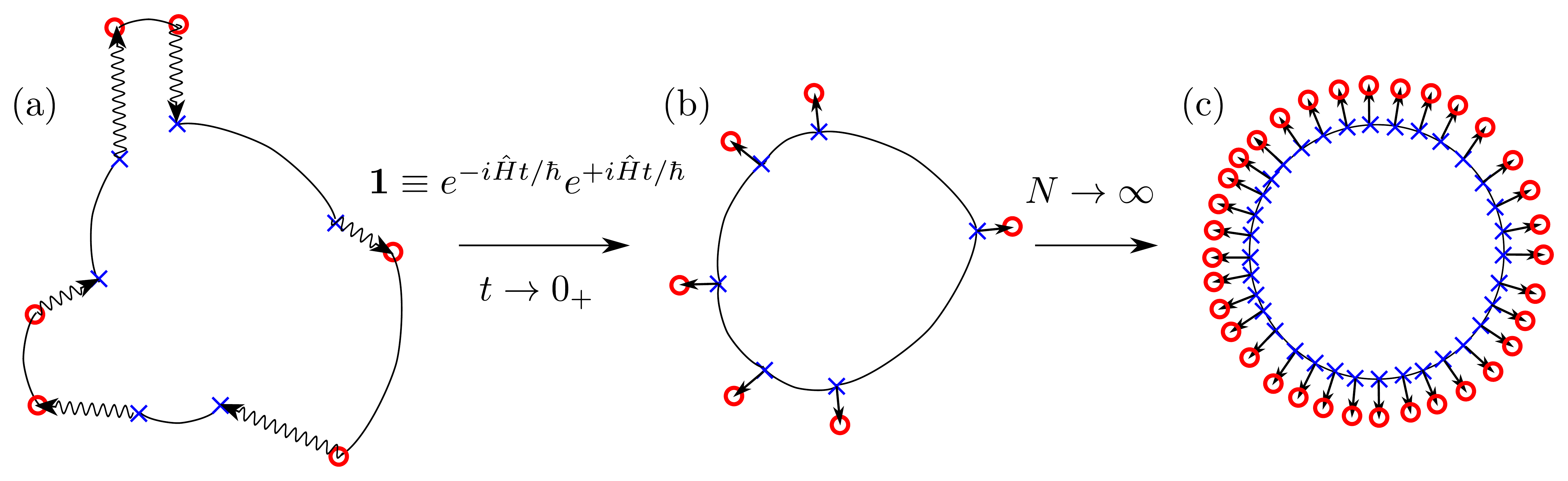}}
\caption{Diagrams showing
 (a) the generalized flux-side time-correlation function  $\Cfsun{t}$ of Eq.~\eqref{eq:gencfs};
  (b) the \shortt~limit of $\Cfsxv{t}$, Eq.~\eqref{eq:likenf2}; 
  (c)  the latter for a large value
  of $N$. Sinusoidal lines represent real-time evolution, curved lines imaginary-time evolution, and the symbols indicate the places  acted on by the
   flux operator $\mathcal{\hat F}[\fq]$ (blue crosses) and the side operator $h[\gz]$ (red circles).}
\end{figure*}

\subsection{Positive-definite Boltzmann statistics}

Having shown that the \shortt~limit of \eqr{gencfs} is non-zero if $\fq \equiv \gq$, we now determine the conditions on $\fq$ that give rise to
 positive-definite quantum statistics. The special case $\xi_i^-=1/N,\xi_i^+=0$ has already been treated in Paper I and we use the same approach here for the more general case, which is to find the condition on $\fq$ which guarantees that the integral over $\tdo$ in \eqr{likenf2} is positive in the limit \largeN. We first express the Boltzmann operator in ring polymer form,
\begin{align}
\lim_{N\to\infty} & \prod_{j=0}^{2N-1} \bra{Q_{j-1}- T_{j-1\ 0} \tilde D_0/2} e^{-\beta \xi_j \hat H} \ket{Q_j+ T_{j0} \tilde D_0/2} \nonumber \\
  = & \prod_{j=0}^{2N-1} \sqrt{\frac{m}{2\pi\beta\xi_j\hbar^2}} \no\\
 & \times e^{-\beta \xi_j[ V(Q_{j-1}- T_{j-1\ 0} \tilde D_0/2) + V(Q_j+ T_{j0} \tilde D_0/2)]/2} \nonumber \\
 & \times e^{-m [ Q_{j} - Q_{j-1} + \tilde D_0(T_{j-1\ 0} + T_{j0})/2]^2/2\beta \xi_j \hbar^2} \label{eq:bbk} 
\end{align}
and note that $T_{j0} \sim N^{-1/2}$, which ensures that the potential energy terms are independent of $\tilde D_0$ in the limit \largeN. \footnote{For equally spaced imaginary-time intervals, the leading non-zero term in the potential goes as $T_{j0}^2 \sim N^{-1}$, giving more rapid convergence with respect to $N$ than the general case of unequally-spaced intervals discussed in the text.}
Expanding the spring term, 
\begin{align}
\lim_{N\to\infty} &  \sum_{j=0}^{2N-1} \frac{m}{2\beta \xi_j \hbar^2} [Q_{j} - Q_{j-1} + \td D_0(T_{j-1\ 0} + T_{j0})/2]^2 \nonumber \\
 & = \lim_{N\to\infty} \sum_{j=0}^{2N-1} m [Q_{j} - Q_{j-1}]^2/2\beta \xi_j \hbar^2 \nonumber \\
 & \qquad + m [Q_{j} - Q_{j-1}]\td D_0(T_{j-1\ 0} + T_{j0})/2\beta \xi_j \hbar^2 \nonumber \\
 & \qquad + m \td D_0^2 (T_{j-1\ 0} + T_{j0})^2/8\beta \xi_j \hbar^2 \label{eq:ttterm}
\end{align}
we see that the integral over the Boltzmann operator is guaranteed to be positive if and only if the cross-terms vanish. In other words the condition
\begin{align}
\lim_{N\to\infty} \sum_{j=0}^{N-1} m [Q_{j} - Q_{j-1}]\tdo(T_{j-1\ 0} + T_{j0})/2\beta \xi_j \hbar^2 = 0 \label{eq:zerocon} 
\end{align}
must be satisfied for the Boltzmann statistics to be positive-definite.
In Appendix~\ref{app:piiti}, we show that this condition
is equivalent to requiring the dividing surface $\fQ$ to be invariant under imaginary-time translation. This was the same conclusion reached in Paper I, starting from the special case of $\xi_i^-=1/N,\xi_i^+=0$.

\subsection{Emergence of RPMD-TST}

When $\fq$ is invariant under imaginary-time translation we can integrate out $\tdo$ and $\tpo$ (see Appendix~\ref{ap:uneq}), to obtain
\begin{align}
  \lttz\lim_{N\to\infty}\Cfsxv{t} = & \int d{\bf Q} \ \dfQ \sqrt{\frac{\ntn}{2\pi m \beta}} \no \\
  & \times \prod_{j=0}^{2N-1} \bra{Q_{j-1}} e^{-\beta\xi_j \hat H} \ket{Q_j} \label{eq:nop} 
\end{align}
with
\begin{align}
 \ntn 
  & =  \lim_{N\to\infty} \sum_{j=0}^{2N-1}\frac{1}{4\xi_j}\left[\ddp{\fQ}{Q_{j-1}} + \ddp{\fQ}{Q_{j}}\right]^2 
\end{align}
The integral in \eqr{nop} is the generalisation of the RPMD-TST integral of \eqr{nopeq} to unequally spaced imaginary time-slices $\xi_j$. Both expressions converge to the same result in the limit \largeN, i.e.
 \begin{align}
  \lttz\lim_{N\to\infty}\Cfsxv{t} &= k_{Q}^\ddag(\beta)\Qrb\nonumber\\
  &\equiv k_{\rm RPMD-TST}^\ddag(\beta)  
\end{align}  
provided that $\fq \equiv \gq$ and that $\fq$ is invariant under imaginary-time translation. In other words, a positive-definite \shortt~limit can arise from the general time-correlation function \eqr{gencfs} only if $\fq$ is invariant under imaginary-time-translation (in the limit \largeN), in which case this limit is identical to that obtained from the simpler time-correlation function Eq.~(31) in Paper~I, namely RPMD-TST.
 
The above derivation can easily be generalized to multi-dimensions, by following the same procedure as that applied
to \eqr{ubfn} in Sec.~V of  Paper I.

\section{Conclusions}
\label{sec:cc}

We have introduced an extremely general quantum flux-side time-correlation function, and found that its \shortt\ limit is non-zero {\em only} when the flux and side dividing surfaces are the same function of path-integral space, and that it gives 
 positive-definite quantum statistics {\em only} when the common dividing surface is invariant
to imaginary-time translation. This \shortt\ limit is identical to the one that was derived in Paper I starting from a simpler form of flux-side time-correlation function (a special case of the function introduced here),
where it was shown to give a true \shortt\ quantum TST which  is identical to RPMD-TST.
 
 We cannot prove that a yet more general flux-side time-correlation function does not exist  (than the one introduced here) which might
  support a {\em different} non-zero \shortt\ limit, which nevertheless gives positive-definite quantum statistics. However, given that the function introduced here includes all known flux-side time-correlation functions as special cases, we think that this is unlikely.
 
This article therefore provides strong evidence (although not conclusive proof) that the quantum TST of Paper I is unique, in the sense that there is no other \shortt~limit which gives a non-zero quantum TST containing positive-definite quantum statistics. In other words, if one wishes to obtain an estimate of the thermal quantum rate by taking the instantaneous flux through a dividing surface, then RPMD-TST cannot be bettered.

\section{Acknowledgements}
TJHH is supported by a PhD studentship from the Engineering and Physical Sciences Research Council.

\appendix
\section{Derivation of the \shortt~limit of \eqr{gencfs}}
\label{ap:st}
\subsection{Coordinate transformation}
\label{ssec:cot}
The coordinate transform used to convert \eqr{nfs} to \eqr{simpt} is 
\begin{align}
 Q_j = & \left\{
 \begin{array}{ll}
  \tfrac{1}{2} \left( q_i + \Delta_i/2 + y_i - \zeta_i/2 \right), & j = 2i \\
  \tfrac{1}{2} \left( q_i - \Delta_i/2 + y_i + \zeta_i/2 \right), & j = 2i +1 
 \end{array}
 \right. \\
 D_j = & \left\{
 \begin{array}{ll}
  - q_i - \Delta_i/2 + y_i - \zeta_i/2, & j = 2i \\
  q_i - \Delta_i/2 - y_i - \zeta_i/2, & j =2i + 1
 \end{array}
 \right. \\
 Z_j = & \left\{
 \begin{array}{ll}
 z_i - \eta_i/2, & j = 2i \\
 z_i + \eta_i/2, & j = 2i + 1
 \end{array}
 \right.
\end{align}
where $j = 0, 1, \ldots, 2N-1$ and $i = 0, 1, \ldots, N-1$. The 
associated Jacobian is unity. Note that $\fq$ is of course unchanged by the coordinate transformation, so $f(\bQ,\bD)$ in \eqr{simpt} depends on $\bQ$ and $\bD$ through the relation
\begin{align}
q_i = & \ Q_{2i} + Q_{2i+1} + (D_{2i+1} - D_{2i})/2, \label{eq:qdef}
\end{align}
i.e.\ $f(\bQ,\bD)$ is {\em not} a general function of $\bQ$ and $\bD$, since it remains a function of only $N$ independent variables. Similarly, $g(\bZ)$ depends on $\bZ$ through
\begin{align}
z_i = & \ (Z_{2i} + Z_{2i+1})/2. \label{eq:zdef}
\end{align}

\subsection{The \shortt~limit}
\label{ssec:stl}
The \shortt~limit of \eqr{simpt} can be obtained by a straightforward application of Eqs.~\eqref{eq:shorttlim}--\eqref{eq:etfop}, and is
\begin{align}
  \lttz & \Cfsun{t} = \lttz\tphtN \int d{\bf Q} \int d {\bf P} \int d {\bf D} \no\\
 \times & \delta[f({\bf Q,D})] S_f({\bf Q,D,P}) h[g({\bf Q + P}t/m)] \nonumber \\
 \times & \prod_{j=0}^{2N-1} \bra{Q_{j-1}-D_{j-1}/2} e^{-\beta \xi_j \hat H} \ket{Q_j+D_j/2} e^{i D_j P_j/\hbar} \label{eq:stl} 
\end{align}
where $P_j = (Z_j - Q_j)m/t$, and
\begin{align}
 S_f & ({\bf Q,D,P}) = \frac{1}{2m} \smiN \ddp{\fq}{q_i} p_i 
\end{align}
\begin{align}
= & \frac{1}{2m} \smiN \ddp{f({\bf Q,D})}{[Q_{2i} + Q_{2i+1} + (D_{2i+1} - D_{2i})/2]} \nonumber \\
  & \times \left[P_{2i}+P_{2i+1} + \frac{m}{2t}(D_{2i+1} - D_{2i}) \right]
\end{align}
with $p_i = (z_i-q_i)m/t$. 

To convert \eqr{stl} to \eqr{bigeq}, we note that
\begin{align}
 \ddp{g({\bf Z})}{Z_{2i}} = \ddp{g({\bf Z})}{Z_{2i+1}},
\end{align}
[see \eqref{eq:zdef}] and hence that
\begin{align}
 \lim_{t\to0_+} g(\bQ+\bP t/m) = & g({\bf Q}) + \frac{t}{m} \sum_{i=0}^{N-1} (P_{2i} + P_{2i+1}) \ddp{g({\bf Q})}{Q_{2i}}.
\end{align}
Transforming to
\begin{align}
 P^+_i = & \ \ort (P_{2i}+P_{2i+1}) \\
 P^-_i = & \ \ort (P_{2i} - P_{2i+1})
\end{align}
where $0 \leq i \leq N-1 $ and likewise for $\bf D^+,D^-$, we obtain
\begin{widetext}
\begin{align}
  \lttz\Cfsun{t} = & \lttz\tphtN \int d{\bf Q} \int d {\bf P^+} \int d {\bf P^-} \int d {\bf D^+} \int d {\bf D^-} \delta[f({\bf Q,D^-})] S_f({\bf Q,D^-,P^+}) h[g(\bQ+\sqrt{2}\bP^+ t/m)] \nonumber \\
  & \times \prod_{i=0}^{N-1} \Big[ e^{i D^+_i P^+_i/\hbar} e^{i D^-_i P^-_i/\hbar} \bra{Q_{2i-1}- \otrt(D_{i-1}^+ - D^-_{i-1})} e^{-\beta \xi_{2i} \hat H} \ket{Q_{2i}+\otrt(D_i^+ + D_i^-)} \nonumber \\
  & \qquad\times \bra{Q_{2i} - \otrt(D_i^+ + D_i^-)} e^{-\beta \xi_{2i+1} \hat H} \ket{Q_{2i+1} + \otrt(D_i^+ - D_i^-)}\Big].\label{eq:bigeq}
\end{align}
\end{widetext}
We can then integrate out the $\bf P^-$ to generate $N$ Dirac delta functions in $\bf D^-$, such that $f({\bf Q,D^-})$ and $S_f({\bf Q,D^-,P^+})$ reduce to
 $\fQ$ and $S_f({\bf Q,P^+})$, and \eqr{bigeq} becomes
\begin{align}
  \lttz & \Cfsun{t} = \lttz\tphN \int d{\bf Q} \int d {\bf P^+} \int d {\bf D^+} \no\\
  & \times \dfQ S_f({\bf Q,P^+}) h[g(\bQ+\sqrt{2}\bP^+ t/m)] \nonumber \\
  & \times \prod_{i=0}^{N-1} \bra{Q_{2i-1}-\otrt D_{i-1}^+} e^{-\beta \xi_{2i} \hat H} \ket{Q_{2i}+\otrt D_i^+} \nonumber \\
  & \qquad \times \bra{Q_{2i}-\otrt D_i^+} e^{-\beta \xi_{2i+1} \hat H} \ket{Q_{2i+1}+\otrt D_i^+}\no \\
  & \qquad \times e^{i D^+_i P^+_i/\hbar}
\end{align}
It is easy to show (following the reasoning given in Sec.~IIIB of Paper I) that this expression is non-zero only if
 $\fQ \equiv \gQ$, in which case the limit
\begin{align}
\lim_{t \to 0_+} & \dfQ h[f(\bQ+\sqrt{2}\bP^+ t/m)] \no\\
& = \lim_{t \to 0_+} \dfQ h[f({\bf Q}) + t S_f({\bf Q,P^+})] \nonumber \\
& = \dfQ h[S_f({\bf Q,P^+})]
\end{align}
 results in \eqr{nzq}.

\subsection{Normal mode transformation}
\label{ssec:intout}

To integrate out $D_{i}^+,\ i>0$ from \eqr{nzq}, we transform to the coordinates 
\begin{align}
 \tilde P_j' = & \sum_{i=0}^{N-1} P_i^+ T_{2ij}' 
\end{align}
\begin{align}
 \tilde D_j' = & \sum_{i=0}^{N-1} D_i^+ T_{2ij}'
\end{align}
where
\begin{align}
 T_{i0}' = & \frac{1}{\sqrt{B_N}}\ddp{f({\bf Q})}{Q_{i}}, \label{eq:tdef} 
\end{align}
\begin{align}
 B_N' = & \sum_{i=0}^{N-1} \left[ \ddp{f({\bf Q})}{Q_{2i}} \right]^2 \label{eq:bnpdef} 
\end{align}
such that $S_f({\bf Q,P^+}) = \tpo'\sqrt{2B_N'}$ and, from \eqr{qdef}, $T_{2i0}' = T_{2i+1 0}'$. The other normal modes, $T_{ij}'$, $j = 1, \ldots, N-1$ are chosen to be orthogonal to $T_{i0}'$ and their exact form need not concern us further. Unless $\fQ$ is linear in $\bQ$ (such as a centroid), $T_{ij}'$ and $B_N$ are functions of $\bQ$. We obtain
\begin{widetext}
\begin{align}
 \lttz \Cfsxv{t} = & \tphN \int d{\bf Q} \int d {\bf \tilde P'} \int d {\bf \tilde D'} \ h(\tilde P_0') \frac{\tpo'}{m} \sqrt{B_N'} \dfQ \prod_{i=0}^{N-1} e^{i \tilde D_i' \tilde P_i'/\hbar} \nonumber \\
 \times & \prod_{j=0}^{2N-1} \bra{Q_{j-1}-\otrt \sum_{i=0}^{N-1} T_{j-1\ i}' \tilde D_{i}'} e^{-\beta \xi_j \hat H} \ket{Q_j+ \otrt \sum_{i=0}^{N-1} T_{ji}' \tilde D_{i}'}
\end{align}
\end{widetext}
Integrating out $\tilde P_i', \ 1 \leq i \leq N-1$ to generate Dirac delta functions in $\tilde D_i', \ 1 \leq i \leq N-1$, which are themselves then integrated out, we obtain
\begin{align}
 & \lttz \Cfsxv{t} = \frac{1}{2\pi\hbar} \int d{\bf Q} \int d \tilde P_0' \int d \tilde D_0' \no\\
 & \times h[\tilde P_0]\frac{\tilde P_0'}{m}  \sqrt{2B_N} \dfQ e^{i \tilde D_0' \tilde P_0'/\hbar} \nonumber \\
 & \times \prod_{j=0}^{2N-1} \bra{Q_{j-1}- \otrt T_{j-1\ 0}' \tilde D_0'} e^{-\beta \xi_j \hat H} \ket{Q_j+  \otrt T_{j0}' \tilde D_0'}.
 \label{eq:pdp} 
\end{align}
This transformation was made using the $N$-dimensional $\bf P^+, D^+$ coordinates. To redefine the transformation from $2N$-dimensional $\bP, \bD$ we define $\bf \tilde P\ \tilde D$ (where the absence of a prime indicates a $2N$-dimensional transformation), such that [using \eqr{qdef}]
\begin{align}
 \tilde P_0' & = \frac{\sum_{i=0}^{N-1}P_i^+\ddp{\fQ}{Q_{2i}}}{\sqrt{\sum_{i=0}^{N-1}\left(\ddp{\fQ}{Q_{2i}}\right)^2}} \\
  & =  \frac{\sum_{i=0}^{2N-1}P_i \ddp{\fQ}{Q_{i}}}{\sqrt{\sum_{i=0}^{2N-1}\left(\ddp{\fQ}{Q_{i}}\right)^2}} \\
  & = \tpo.
\end{align}
Likewise $\tilde D_0' = \tdo$. However, from \eqr{bnpdef} 
\begin{align}
 B_N' & = \frac{1}{2} \sum_{i=0}^{2N-1} \left[ \ddp{f({\bf Q})}{Q_{i}} \right]^2 \\
 & = \frac{1}{2} B_N
\end{align}
and it follows from this result \eqr{tdef} that $T_{j0} = T_{j0}'/\sqrt{2}$.
These adjustments convert \eqr{pdp} to Eq.~\eqref{eq:likenf2}.
 
\section{Invariance of the dividing surface to imaginary-time translation}
\label{app:piiti}
To show that Eq.~\eqref{eq:zerocon} is equivalent to the requirement that $\fq$ be invariant under imaginary time-translation (in the limit \largeN),
we rewrite this expression in the form
\begin{align}
\lim_{N\to\infty} \sum_{j=0}^{2N-1} T_{j0} \left(\frac{Q_{j+1}-Q_j}{\beta \hbar \xi_{j+1}}-\frac{Q_{j-1} - Q_{j}}{\beta \hbar\xi_j}\right) = 0 \label{eq:tqc}. 
\end{align}
We then consider a shift in the imaginary-time origin by a small, positive, amount $\delta \tau$,
 which we represent by the operator $\mathcal{P_{+\delta \tau}}$. We then obtain
\begin{equation} 
\lim_{N\to\infty} \mathcal{P_{+\delta \tau}}Q_j=Q_j+(Q_{j+1}-Q_j)\delta \tau/\xi_{j+1}
 \end{equation}
and hence 
\begin{align}
 \lim_{N\to\infty} & \mathcal{P_{+\delta \tau}} \fQ \no \\ 
 & =  \lim_{N\to\infty}\fQ +  \sum_{j=0}^{2N-1} (Q_{j+1}-Q_j) \ddp{\fQ}{Q_j} \frac{\delta \tau}{\beta \hbar \xi_{j+1}}, \label{eq:permdef}
\end{align}
Noting from \eqr{tdef} that $\partial \fQ/\partial Q_j = \sqrt{B_N}T_{j0}$, we see that
 the second term on the RHS of \eqr{permdef} is proportional to the first term on the LHS of \eqr{tqc}. Using similar reasoning, we find that the second term on the  LHS of \eqr{tqc} is proportional to $-\lim_{N\to\infty} \mathcal{P_{-\delta \tau}} \fQ$, where $\mathcal{P_{-\delta \tau}}$ denotes a shift in the imaginary-time origin by a small, negative, amount $-\delta \tau$. \eqr{tqc} is thus equivalent to the condition
\begin{align}
\lim_{N\to\infty}\mathcal{P_{+\delta \tau}}\fQ - \mathcal{P_{-\delta \tau}} \fQ = 0,
\end{align}
i.e.\ that the dividing surface $\fQ$ is invariant to imaginary-time-translation in the limit \largeN.

\section{Integrating out the ring-opening coordinate}
\label{ap:uneq}

When Eq.~(\ref{eq:zerocon}) is satisfied, the only contribution to the imaginary-time path-integral from $D_0$ in the limit \largeN\ is the term ${m \tdo^2}A({\bf Q})/{2\beta\hbar^2}$, in which
\begin{align}
A({\bf Q}) &= \lim_{N\to\infty} \sum_{j=0}^{2N-1} \frac{1}{4\xi_j} \left[T_{j-1\ 0} + T_{j0}\right]^2 \\
&= \lim_{N\to\infty}\frac{1}{B_N}\sum_{j=0}^{2N-1}\frac{1}{4\xi_j}\left[\ddp{\fQ}{Q_{j-1}} + \ddp{\fQ}{Q_{j}}\right]^2
\end{align}
and where the last line follows from the definition of $T_{j0}$ in Appendix A. The integral over $\tdo$ in \eqr{likenf2} is then easily evaluated to give
\begin{align}
  \lttz & \Cfsxv{t} = \frac{1}{2\pi\hbar} \int d{\bf Q} \int d \tilde P_0 \ h[\tilde P_0]\frac{\tilde P_0}{m}  \sqrt{B_N} \dfQ \nonumber \\
  & \times \sqrt{\frac{2\pi\beta \hbar^2}{mA({\bf Q})}}e^{-\beta \tilde P_0^2/2mA({\bf Q})} \prod_{j=0}^{2N-1} \bra{Q_{j-1}} e^{-\beta\xi_j \hat H} \ket{Q_j} \label{eq:uneqd} 
\end{align}
and integration over $\tpo$ gives Eq.~(\ref{eq:nop}).

\bibliography{/home/tjhh2/Projects/articles/refbig}

\end{document}